\documentclass[twocolumn,prl,superscriptaddress,showpacs]{revtex4}
\usepackage{amsthm}
\usepackage{graphicx}
\usepackage{amsmath}
\usepackage{amssymb}
\usepackage{verbatim}
\usepackage{titlesec}
\titlespacing{\section}{0mm}{1.5ex plus .1ex minus .2ex}{5pt}
\titleformat{\section}[runin]{\itshape}{\thesection}{.5em}{}[.]
\titleformat{\subsection}[runin]{\bfseries}{\thesubsection}{.5em}{}[.]
\titleformat{\subsubsection}[runin]{\itshape}{\thesubsubsection}{.5em}{}[.]

\newtheorem{theorem}{Theorem}
\newtheorem{lemma}{Lemma}

\newcommand*{\cH}{\mathcal{H}}

\newcommand{\ket}[1]{|#1\rangle}
\newcommand{\bra}[1]{\langle #1 |}

\newcommand{\proj}[1]{\ket{#1}\bra{#1}}

\newcommand{\ot}[0]{\otimes}

\newcommand{\beq}{\begin{equation}}
\newcommand{\eeq}{\end{equation}}
\newcommand{\best}{\begin{equation*}}
\newcommand{\eest}{\end{equation*}}
\newcommand{\tr}[1]{{\rm Tr}(#1)}
\newcommand{\Tr}{{\rm Tr}}

\newcommand{\rhoAB}{\rho_{AB}}

\newcommand{\ib}[2]{I_b^{(#1)}(#2)}

\newcommand{\ibtot}[2]{(I_b)_{#1}(#2)}

\begin{document}

\title{Broadcast copies reveal the quantumness of correlations}

\author{M. Piani}
\affiliation{Institute for Quantum Computing \& Department of Physics
  and Astronomy, University of Waterloo, 200 University Ave. W., N2L 3G1 Waterloo, Canada}
\author{M. Christandl}
\affiliation{Faculty of Physics, Ludwig-Maximilians-Universit\"at M\"unchen, Theresienstrasse 37, 80333 Munich,
Germany}
\author{C. E. Mora}
\affiliation{Institute for Quantum Computing \& Department of Physics
  and Astronomy, University of Waterloo, 200 University Ave. W., N2L 3G1 Waterloo, Canada}
\author{P. Horodecki}
\affiliation{Faculty of Applied Physics and Mathematics, Gda\'nsk University of Technology, 80-952 Gda\'nsk, Poland}
\affiliation{National Quantum Information Centre of Gda\'nsk, 81-824 Sopot, Poland}

\begin{abstract}
We study the \emph{quantumness} of bipartite correlations by proposing a quantity that combines a measure of total correlations --mutual information-- with the notion of \emph{broadcast copies} --i.e. generally non-factorized copies-- of bipartite states. By analyzing how our quantity increases with the number of broadcast copies, we are able to classify classical, separable, and entangled states. This motivates the definition of the \emph{broadcast regularization of mutual information}, the asymptotic minimal mutual information per broadcast copy, which we show to have many properties of an entanglement measure.
%and is strictly positive for and only for entangled states.
%in particular it is strictly positive for all entangled states ---a signature of the \emph{monogamy of entanglement}.
%Interestingly, when we impose symmetry on the broadcast copies, the correlation measure reduces to a known entanglement measure: the squashed entanglement with classical extensions.
%As such, it is, e.g., potentially useful in tackling problems in computational complexity~\cite{scott}. In particular, $I_b^{(\infty)}$ lies between the classical- and the quantum-conditioned versions of squashed entanglement. While the latter is not know to be strictly positive for all entangled states, we prove that $I_b^{(\infty)}$ is: this fact may be considered a signature of monogamy of entanglement.
\end{abstract}

\maketitle

Much work has recently been performed in order to analyze how correlations can be understood, quantified and classified as either classical or quantum~\cite{recent,localbroad}. Such studies, that in a way go beyond the standard entangled-versus-separable~\cite{revhoro} distinction, are relevant not only for our understanding of the fundamental differences between the classical and quantum world, but also from the point of view of Quantum Information Processing~\cite{NC}.
Indeed, Entanglement is a necessary prerequisite for a task such as Quantum Key Distribution~\cite{Lutken}, but its role in Quantum Computation is less clear, as
there are cases where quantum correlations that are weaker than entanglement seem to be sufficient to boost performance with respect to classical computation~\cite{DQC1}.

In studying the \emph{quantumness} of correlations, researchers have focused on the following hierarchy of states. \emph{Classical-classical} \mbox{(CC-)} states are of the form $\sum_{ij}p^{AB}_{ij}|i\rangle\langle i|_A\otimes |j\rangle\langle j|_B$, with $\{\ket{i}_A\}$ and $\{\ket{j}_B\}$ orthonormal bases, and $\{p^{AB}_{ij}\}$ a joint probability distribution. A CC-state is the embedding of a probability distribution in the formalism of quantum theory and as such has no
quantumness. CC-states are a proper subclass the class of \emph{separable states}, which are of the form $\sum_{k}p_{k}\rho^k_A\otimes \rho^k_B$ for a probability distribution $\{p_k\}$ and local quantum states $\rho_A^k$ and $\rho_B^k$. Separable states can be generated with Local Operations and Classical Communication (LOCC) only and are therefore considered to have little quantumness. The remaining states are called \emph{entangled} and exhibit the most quantumness.

In this Letter we study the quantumness of correlations by combining a measure of total correlations ---mutual information (MI)--- and the notion of \emph{broadcast copies}, i.e.~generally non-factorized copies (see Figure~\ref{fig:broadcastcond}) of a bipartite quantum state.
We relate quantumness to monogamy of correlations, and in particular to \emph{monogamy of entanglement}, which in standard terms refers to the impossibility of a system to be strongly entangled with two or more other systems at the same time~\cite{CKW}. Here, we adopt a different perspective by considering broadcast copies, and analyze quantitatively the minimal growth of the correlations with the number of broadcast copies. Whereas for factorized copies the correlations increase linearly for all states, this is not
%necessarily
true for non-factorized copies. Indeed, for CC-states correlations do not have to increase at all and can be freely shared among any number of broadcast copies. We show that for non-CC separable states there is actually an increase, but it is bounded, while for entangled states the correlations \emph{must increase linearly} with the copies, a result we term \emph{copy-copy monogamy of entanglement}. This is better expressed in quantitative terms by introducing the \emph{broadcast regularization of MI},
%defined as
the minimal per-copy MI between parties, when they share an infinite amount of broadcast copies. We show that this quantity has many properties of an entanglement measure~\cite{pleniorev}, we establish relations with known entanglement measure, and we conjecture that it is an entanglement measure itself. We then restrict the minimization to permutationally-invariant broadcast copies and prove that the corresponding constrained broadcast regularization of MI equals the classical version of squashed entanglement~\cite{multisquash}.

\section{Broadcast copies and mutual information}

\begin{figure}
\includegraphics[width=0.4\textwidth]{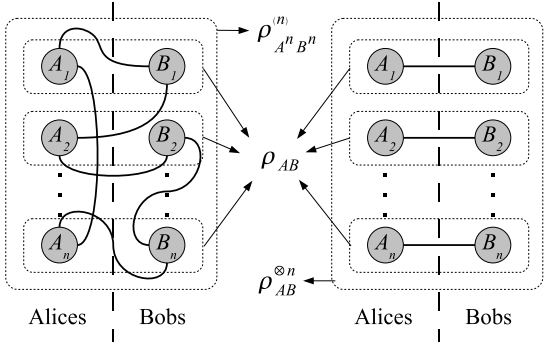}
\caption{$n$ copies of a bipartite state $\rho_{AB}$: broadcast copies (left) and factorized copies (right). Solid lines connecting subsystems symbolize correlations. The vertical dashed line indicates the bipartite cut across which correlations are quantified by mutual information.}
\label{fig:broadcastcond}
\end{figure}

In~\cite{localbroad} the quantumness of correlations of a bipartite state $\rho_{AB}$ on Hilbert space $\cH_A\otimes\cH_B$ was addressed from an operational point of view by employing the notion of broadcast copies and by quantifying total correlations by means of MI. The MI of a state $\rho\equiv\rho_{AB}$ is defined as $I(\rho)\equiv I(A:B)_{\rho}\equiv S(A)_\rho+S(B)_\rho-S(AB)_\rho$, with $S(X)_{\rho}=-\Tr\rho_X\log_2\rho_X$ the von Neumann entropy of subsystem $X$ when the state of the total system is $\rho$~\footnote{MI is a fundamental measure of correlations (see~\cite{recent} and references therein) that: does not increase under local operations; is neither convex nor concave, but respects $I(\sum_kp_k\rhoAB^k)\leq\sum_kp_kI(\rhoAB^k)+S(\{p_k\})$, with $S(\{p_k\})\equiv-\sum_k p_k \log_2 p_k$ the Shannon  entropy of the classical probability distribution $\{p_k\}$; is additive: $I(\rho_{A_1B_1}\otimes\sigma_{A_2B_2})=I(\rho_{A_1B_1})+I(\sigma_{A_2B_2})$.}.
We say that a state $\rho^{(n)}_{X^n}$, $X^n\equiv X_1\ldots X_n$, is an \emph{$n$-copy broadcast state of $\rho$} if $\rho^{(n)}_{X_k}\equiv {\rm Tr}_{X_1 \cdots X_{k-1} X_{k+1} \cdots X_n}\rho^{(n)}=\rho$ for all $k$. Each system $X_k$ may be composed of subsystems, in our case $X_k=A_kB_k$.
%As illustrated in Figure~\ref{fig:broadcastcond},
Broadcast copies may contain correlations among the different copies, in contrast to factorized copies $\rho_X^{\otimes n}$ (Figure~\ref{fig:broadcastcond}).
For example, given any mixed ensemble $\{(p_k,\rho^k_{AB})\}$ for $\rho_{AB}$, i.e.~$\rho_{AB}=\sum_k p_k \rho^k_{AB}$, the convex combination of factorized states $\rho^{(n)}[\{(p_k,\rho^k_{AB})\}]\equiv\sum_k p_k {\rho^k_{AB}}^{\otimes n}$ is a possible $n$-copy broadcast state of $\rho_{AB}$. Such states are also known as \emph{de Finetti states} and play an important role in quantum versions of de Finetti's theorem~\cite{deFin, definettirenner}. By combining MI and the notion of broadcast copies, we can define the \emph{$n$-copy broadcast MI} of $\rho_{AB}$ as $$\ibtot{n}{\rho_{AB}}\equiv\min_{\rho^{(n)}}I(A^n:B^n)_{\rho^{(n)}},$$
where the minimum is taken over all $n$-copy broadcast states $\rho^{(n)}_{A^nB^n}$ of $\rho_{AB}$. A broadcast copy $\bar{\rho}^{(n)}$ such that $I(\bar{\rho}^{(n)})=\ibtot{n}{\rho_{AB}}$ will be said to be optimal~\footnote{As the set of $n$-copy broadcast states is compact, there exist optimal broadcast copies.}.
In \cite{localbroad} a \emph{no-local-broadcasting theorem} for quantum correlations was derived by proving that for non-CC states---even separable ones---one has $\ibtot{n}{\rho_{AB}}>I(\rho_{AB})$, for $n\geq2$. This suggests that the quantumness of the correlations
present in
$\rho_{AB}$
may be revealed by the dependence of $\ibtot{n}{\rho_{AB}}$ on the number of broadcast copies $n$~\footnote{Notice that $I(\rho_{AB})\leq\ibtot{n}{\rho_{AB}}\leq I(\rho_{AB}^{\otimes n})=n I(\rho_{AB})$.}.
We will particularly focus on its behaviour for large $n$, as given by
the \emph{broadcast regularization of MI}~\footnote{The broadcast regularization of any a state-dependent real function $f$ will be denoted by $f^{(\infty)}_b$. The standard regularization $f^{(\infty)}$
is defined as $f^{(\infty)}(\rhoAB)\equiv\lim_n \frac{1}{n}f(\rho_{AB}^{\otimes n})$, i.e. on factorized copies. For MI, one has trivially $I^{(\infty)}=I$.}
\[
\ib{\infty}{\rho_{AB}}\equiv\lim_n \frac{1}{n}\min_{\rho^{(n)}}I(A^n:B^n)_{\rho^{(n)}}.
\]
In the following theorem we formalize the intuition that classical correlations can be freely shared among the broadcast copies, while quantum correlations can not.
\begin{theorem}
\label{thm:main}
The $n$-copy broadcast MI $(I_b)_n$ as a function of $n$: (i) is constant for CC-states; (ii) grows (strictly from one to two copies) but is bounded for separable states that are not CC-states; (iii) grows strictly and asymptotically linearly for all entangled states.
\end{theorem}
\begin{proof}
(ii) By definition, given any mixed ensemble realization $\{(p_k,\rho^k_{AB})\}$ of $\rho_{AB}$, we have $\ibtot{n}{\rho_{AB}}\leq I\big(\rho^{(n)}[\{(p_k,\rho^k_{AB})\}]\big)$. For separable states one may choose an ensemble with $\rho^k_{AB}=\rho^k_A\otimes \rho^k_B$, for all $k$. Then, independently of the number of copies $n$, $I\big(\rho^{(n)}[\{(p_k,\rho^k_A\otimes \rho^k_B)\}]\big)\leq S(\{p_k\})$. In the separable non-CC case, the strict growth from $n=1$ to $n=2$ of $(I_b)_n$ was proved in~\cite{localbroad}. (i) For CC-states, one can relabel $k=(i,j) $ and set $\rho^k_{AB}\equiv|i_A\rangle\langle i_A|\otimes |j_B\rangle\langle j_B|$, so that $I(\rho_{AB})=\ibtot{n}{\rho_{AB}}=I\big(\rho^{(n)}[\{(p_{ij},|i_A\rangle\langle i_A|\otimes |j_B\rangle\langle j_B|)\}]\big)=I(\{p_{ij}\})$~\footnote{Here, $I(\{p^{AB}_{ij}\})\equiv S(\{p^{A}_{i}\})+S(\{p^{B}_{j}\})-S(\{p^{AB}_{ij}\})\equiv S(\{p^{AB}_{ij}\}\|\{p^{A}_{i}p^{B}_{j}\})$ is the classical MI of the joint probability distribution $\{p^{AB}_{ij}\}$ with marginal distributions $p^A_i=\sum_jp^{AB}_{ij}$ (similarly for $B$).
$S(\{p_k\}\|\{q_k\})\equiv -\sum_k p_k \log_2 \frac{q_k}{p_k}$ is the \emph{Kullback-Leibler distance} between two probability distributions $\{p_k\}$ and $\{q_k\}$.} which is a constant independent of $n$.
(iii) By definition, $I^{(\infty)}_b=\lim_n\frac{1}{n}(I_b)_n$, therefore $(I_b)_n\geq n I^{(\infty)}_b$~\footnote{Take an optimal broadcast copy $\bar{\rho}^{(n)}$. Then $(I_b)_n(\rho)=n\frac{I((\bar{\rho}^{(n)})^{\otimes k})}{nk}\geq n \frac{(I_b)_{nk}(\rho)}{nk}$, and taking the limit $k\rightarrow\infty$ proves the claim.}. The claim follows then from the statement that $I^{(\infty)}_b$ is strictly positive for all entangled states (Lemma~\ref{lem:strict} below).
\end{proof}

\section{Properties of $I^{(\infty)}_b$}
The next theorem establishes many of the properties of $I^{(\infty)}_b$.
\begin{theorem}
\label{thm:properties}
The broadcast regularization of MI $I^{(\infty)}_b$ is:
(i) zero for separable states; (ii) convex; (iii) monotone under local operations: $I((\Lambda_A\otimes\Lambda_B)[\rhoAB])\leq I(\rhoAB)$, for completely-positive trace-preserving maps $\Lambda_A$ and $\Lambda_B$;
(iv) subadditive: $I^{(\infty)}_b(\rho_{AB}\otimes\sigma_{A'B'})\leq I^{(\infty)}_b(\rho_{AB})+I^{(\infty)}_b(\sigma_{A'B'})$; (v) weakly additive: $I^{(\infty)}_b(\rho_{AB}^{\otimes m})=mI^{(\infty)}_b(\rho_{AB})$; (vi) asymptotically continuous: for $\epsilon\equiv||\rho_{AB}-\sigma_{AB}||_1<(\frac{2}{21})^2$, $||X||_1=\Tr\sqrt{X^\dagger X}$, one has $|I^{(\infty)}_b(\rho_{AB})-I^{(\infty)}_b(\sigma_{A'B'})| \leq 126\sqrt{\epsilon}\log_2 d+6h(\frac{21}{2}\sqrt{\epsilon})$,
with $h(x)=-x\log_2x-(1-x)\log_2(1-x)$
and $d$ the dimension of $AB$.
\end{theorem}
\begin{proof}
(i) is a consequence of Theorem \ref{thm:main}. (ii) is proved by noting that for optimal broadcast copies $\bar{\rho}_i^{(n)}$ of $\rho_i$: $\ibtot{n}{\sum_ip_i\rho_i}\leq I(\sum_ip_i\rho_i^{(n)})\leq \sum_ip_i\ibtot{n}{\rho_i}+S(\{p_i\})$. (iii) derives from the fact that if $\bar{\rho}^{(n)}$ is an optimal broadcast copy for $\rho$, then $(\Lambda_A^{\otimes n}\otimes\Lambda_B^{\otimes n})[\bar{\rho}^{(n)}]$ is a broadcast copy of $(\Lambda_A \otimes \Lambda_B)[\rho_{AB}]$ and $\ibtot{n}{(\Lambda_A \otimes \Lambda_B)[\rho_{AB}]}\leq I((\Lambda_A^{\otimes n}\otimes\Lambda_B^{\otimes n})[\bar{\rho}^{(n)}])\leq I(\bar{\rho}^{(n)})=\ibtot{n}{\rho}$. (iv) follows from the additivity of MI: $\ibtot{n}{\rho\otimes\sigma}\leq I(\bar{\rho}^{(n)}\otimes\bar{\sigma}^{(n)})=\ibtot{n}{\rho}+\ibtot{n}{\sigma}$, for $\bar{\rho}^{(n)}$ and $\bar{\sigma}^{(n)}$ optimal broadcast copies of $\rho$ and $\sigma$, respectively. Given subadditivity, in order to prove (v) it is sufficient to observe that
$I^{(\infty)}_b(\rho^{\ot k})
=\lim_m\frac{1}{m}\ibtot{m}{\rho^{\ot k}}
\geq k\lim_m \frac{1}{mk}\ibtot{mk}{\rho}
%=k \lim_l\frac{1}{l}\ibtot{l}{\rho}
=k I^{(\infty)}_b(\rho)$. The proof of (vi) is relatively technical and will be reported elsewhere~\cite{longAmutual information}. The main idea is to first prove that for any $\rho_X$, $\sigma_X$, $\epsilon\equiv\|\rho_X-\sigma_X\|_1<1$, there exist a quantum operation $\Lambda_X\equiv\Lambda_X(\rho,\sigma)$ such that: (a) $\Lambda_X[\rho_X]=\sigma_X$, and (b) for any extension $\tau_{XY}$ satisfying $\tau_X=\rho_X$, $\|(\Lambda_X\otimes\openone_Y)[\tau_{XY}]-\tau_{XY}\|\leq \frac{21}{2}\sqrt{\epsilon}$. Thus, if $\rho^{(n)}$ is a broadcast copy of $\rho$, there exist $\Lambda$ such that $\sigma^{(n)}=\Lambda^{\otimes n}[\rho^{(n)}]$ is a broadcast copy of $\sigma$ with comparable MI.
\end{proof}

According to Theorem~\ref{thm:properties}, $I^{(\infty)}_b$ has many of the properties of an \emph{entanglement measure}~\cite{pleniorev}, and we conjecture that it really is an \emph{entanglement monotone}, i.e. that it decreases (on average) under LOCC.

\section{Relation to entanglement measures}
A way to prove that $I^{(\infty)}_b>0$ for all entangled states is suggested by noting the relation of $I^{(\infty)}_b$ to known entanglement measures
~\cite{squash,multisquash,squashedoppenheim,condent}:
%\begin{itemize}
%\item
\emph{squashed entanglement} $E^Q_{\rm sq}(\rho_{AB})\equiv \frac{1}{2}\inf_{\rho_{ABE}} \big(I(A:BE)_\rho-I(A:E))_\rho\big)$,
where the infimum is over all \emph{extensions} $\rho_{ABE}$ of $\rho_{AB}$, i.e. states $\rho_{ABE}$ satisfying $\Tr_E(\rho_{ABE})=\rho_{AB}$;
%\item
\emph{conditional entanglement of MI (CEMI)}
$E_I(\rho_{AB}) \equiv \frac{1}{2}\inf_{\rho_{ABA'B'}}\big(I(AA':BB')_\rho-I(A':B'))_\rho\big)$,
with the infimum over extensions of $\rho_{AB}$;
%\item
\emph{classical squashed entanglement}
$E^C_{\rm sq}(\rho_{AB})\equiv\frac{1}{2}\min_{\rho_{AB\hat{E}}} \big(I(A:B\hat{E})_\rho-I(A:\hat{E}))_\rho\big)$,
where the minimum is over all extensions $\rho_{AB\hat{E}}$ of $\rho_{AB}$ that are classical on $\hat{E}$, i.e. $\rho_{AB\hat{E}}=\sum_k p_k\rho^k_{AB}\otimes\proj{k}_{\hat{E}}$.
%\end{itemize}
Squashed entanglement and CEMI obey $E_I\geq E^Q_{\rm sq}$, and have an operational interpretation as minimal quantum communication costs
in quantum state redistribution~\cite{squashedoppenheim,condent}.

By the definition of $E_I(\rho_{AB})$, it holds $I(AA':BB')_{\rho}\geq 2E_I (\rho_{AB})+I(A':B')_{\rho}$ for any extension $\rho_{AA'BB'}$ of $\rho_{AB}$. Therefore, given an $n$-copy broadcast state $\rho_{AB}^{(n)}$, by using recursively the broadcast condition one obtains $\ibtot{n}{\rho_{AB}}\geq 2nE_I(\rho_{AB})$. By dividing both sides of this inequality by $n$ and taking the limit $n\rightarrow\infty$, we get $I^{(\infty)}_b\geq 2E_I$. Nonetheless, neither $E_I$ nor $E^Q_{\rm sq}$ are known to be strictly positive for all entangled states, in particular because the extending systems in the definitions may have any dimension. Interestingly, thanks to the classicality of the extension, $E^C_{\rm sq}$ has a finite-dimensional optimal extension $\hat{E}$ and is thus known to be strictly positive for all entangled states~\cite{multisquash}.

In order to find good lower bounds on $I^{(\infty)}_b$ we consider the classical MI associated to
a bipartite state quantum $\rho_{AB}$~\cite{entpur}, defined as
$I_C(\rho_{AB})\equiv\max_{\{M_i\otimes N_j\}} I(\{p_{ij}(\rho_{AB})\})$.
 The maximum is taken with respect to all local POVMs $M_i\geq 0$, $\sum_i M_i = \openone$ (acting on system $A$) and $N_j\geq 0$, $\sum_i N_j = \openone$, (acting on system $B$) respectively, and
%$I(\{q_{ij}\})$ corresponds to the classical mutual information of the joint probability disribution $\{q_{ij}\}_{ij}$;
$p_{ij}(\rho)=\tr{M_i\otimes N_j \rho}$.
%The extraction of classical correlations from $\rhoAB$ can be interpreted as the transformation of $\rhoAB$ into a CC-state via local measurement operations $\mathcal{M}(X)=\sum_i\Tr(XM_i)\proj{i}$, with $\{M_i\}$ a POVM measurement and $\{\ket{i}\}$ an orthonormal basis.
As MI decreases under local measurements, $I_C(\rho_{AB})\leq I(\rho_{AB})$, with equality if and only if the state $\rhoAB$ is CC~\cite{localbroad}. We now define a quantity in the same spirit of $E_I$:
\[
E_{I_C}(\rho_{AB})\equiv\inf_{\rho}\big(I_C(AA':BB')_\rho-I_C(A':B'))_\rho\big),
\]
with $\rho\equiv\rho_{ABA'B'}$ an extension of $\rho_{AB}$.
%If $I_C(\rho_{AB})$ is interpreted as the amount of classical correlations contained in the quantum state $\rho_{AB}$,
$E_{I_C}$ measures the minimal increase in classical correlations due to ``adding'' two systems $AB$ in the state $\rho_{AB}$ to arbitrary ancillas $A'B'$.
%~\footnote{This result is reminiscent of the findings in~\cite{acinprivacy}.}. 
The following lemma proves that $E_{I_C}(\rho_{AB})>0$ if and only if $\rho_{AB}$ is entangled: \emph{entanglement and only entanglement implies a higher amount of classical correlations}. Furthermore, it relates $I^{(\infty)}_b$ and $E_{I_C}$ and completes the proof of Theorem~\ref{thm:main}.
%~\footnote{Note that, even if $I_C\leq I$, the relation $E_{I_C}\leq E_{I}$ might well be false because of the presence of two competing terms in the definitions of $E_{I}$ and $E_{I_C}$.}.
\begin{lemma}
\label{lem:strict}
It holds that (i) $I^{(\infty)}_b\geq E_{I_C}$, and that (ii) $E_{I_C}$ vanishes for and only for separable states.
\end{lemma}
\begin{proof}
(i) For any $n$-copy broadcast state $\rho^{(n)}$ of $\rho$, we have $I(\rho^{(n)})\geq I_C(\rho^{(n)})\geq nE_{I_C}(\rho)$, where we used again the broadcast condition and the definition of $E_{I_C}$. Thus, $I^{(\infty)}_b\geq(I_C)^{(\infty)}_b\geq E_{I_C}$. (ii) The latter relations prove that $E_{I_C}$ vanishes for separable states.
%\footnote{Another way to see it, given a separable state $\sum_{k}p_{k}\rho^k_A\otimes \rho^k_B$, is to consider an extension $\sigma_{ABA'B'}=\sum_{k}p_{k}\rho^k_A\otimes \proj{k}_{A'}\otimes \rho^k_B\otimes \proj{k}_{B'}$.}.
In order to prove strict positivity on entangled states, consider any extension $\rho_{ABA'B'}$ of a state $\rho_{AB}$. The optimal local measurements for $I_C(\rho_{ABA'B'})$ in general act jointly on $AA'$ and $BB'$. Let us restrict ourselves to measurements $\{M'_k\}$ and $\{N'_l\}$ on $A'$ and $B'$ that attain the maximum in $I_C(\rho_{A'B'})$, and optimize solely over POVMs $\{M_i\}$ and $\{N_j\}$ on $A$ and $B$. Thus, as in the first part of the proof of Theorem 3 in~\cite{localbroad}, by using the definition of MI and the concavity of entropy we find
% \begin{split}
% \begin{aligned}
$I_C(\rho_{ABA'B'})-I_C(\rho_{A'B'})\geq\sup_{\{M_i\otimes N_j\}}\sum_{kl} q_{kl}I(\{p_{ij}(\rho^{kl}_{AB})\})$,
% \end{aligned}
% \end{split}
with $q_{kl}\equiv \Tr(\rho_{A'B'} M'_k\otimes N'_l)$ and $\rho^{kl}_{AB}=\Tr_{A'B'}(\rho_{ABA'B'} M'_k\otimes N'_l)/q_{kl}(\rho_{A'B'})$. We now recall that MI can be expressed as relative entropy~[28].
%$I(A:B)_{\rho_{AB}}=S(\rho_{AB}\|\rho_A\otimes\rho_B)$, where $S(\rho\|\tau)=\Tr(\rho\log_2\rho)-\Tr(\rho\log_2\tau)$ is the relative entropy of two states $\rho$ and $\tau$.
Furthermore, $p^{A}_i(\rho^{kl}_{AB})\equiv \sum_j p_{ij}(\rho^{kl}_{AB})=\Tr_A(M_i\rho^{kl}_{A})$, with $\rho^{kl}_{A}=\Tr_B(\rho^{kl}_{AB})$ (similarly for $p^{B}_j(\rho^{kl}_{AB})$). Thus, we find
\begin{align*}
& \sup_{\{M_i\otimes N_j\}} \sum_{kl} q_{kl}I(\{p_{ij}(\rho^{kl}_{AB})\})\\
%&\geq \sup_{\{M_i\otimes N_j\}}S(\{\sum_{kl} q_{kl} p_{ij}(\rho^{kl}_{AB})\}\|\{\sum_{kl}q_{kl} p^A_{i}(\rho^{kl}_{AB})p^B_{j}(\rho^{kl}_{AB})\})\\
%&= \sup_{\{M_i\otimes N_j\}}S(\{ p_{ij}(\sum_{kl} q_{kl}\rho^{kl}_{AB})\} \| \{ p_{ij}(\sum_{kl}q_{kl} \rho^{kl}_{A}\otimes \rho^{kl}_{B})\})\\
&\geq \sup_{\{M_i\otimes N_j\}}S(\{ p_{ij}(\rho_{AB})\} \| \{ p_{ij}(\sum_{kl}q_{kl} \rho^{kl}_{A}\otimes \rho^{kl}_{B})\})\\
&\geq \inf_{\sigma_{AB}\  \text{separable}}\sup_{\{M_i\otimes N_j\}}S(\{ p_{ij}(\rho_{AB})\} \| \{ p_{ij}(\sigma_{AB})\}),
\end{align*}
where we used the joint convexity of relative entropy, the fact that $\sum_{kl} q_{kl} \rho^{kl}_{AB}=\rho_{AB}$, as well as the separability of $\sum_{kl}q_{kl} \rho^{kl}_{A}\otimes \rho^{kl}_{B}$. This lower bound is independent of $\rho_{ABA'B'}$ and is strictly positive for all entangled states, because there exist informationally-complete local POVMs~\cite{SICPOVM} and the relative entropy vanishes only when the two probability distributions are equal.
\end{proof}

The next theorem formalizes the relation of $I^{(\infty)}_b$ with the mentioned entanglement measures.
\begin{theorem}
\label{thm:sequence}
We have the sequence of inequalities:
\beq
\label{eq:sequence}
2E^C_{\rm sq}{\geq} 2\big( E^C_{\rm sq}\big)^{(\infty)}{\geq} I^{(\infty)}_b {\geq} 2E_I {\geq} 2E^Q_{\rm sq}.
\eeq
\end{theorem}
\begin{proof}
The two rightmost inequalities have already been discussed. The first inequality is due to subadditivity of $E^C_{\rm sq}$. The inequality $2E_{\rm sq}^C\geq I_b^{(\infty)}$ is proved by noticing that $E^C_{\rm sq}$ corresponds to $E^C_{\rm sq}(\rho_{AB})=\frac{1}{2}\min_{\{(p_k,\rho^k_{AB})\}}\sum_kp_kI(\rho^k_{AB})$, with the minimum over mixed ensembles for $\rhoAB$~\cite{multisquash}. By choosing an ensemble $\{(\bar{p}_k,\bar{\rho}^k_{AB})\}$  optimal for $E^C_{\rm sq}$, and using additivity of MI, one finds $\ibtot{n}{\rhoAB}\leq I(\rho^{(n)}[\{(\bar{p}_k,\bar{\rho}^k_{AB})\}])\leq n\sum_k\bar{p}_kI(\bar{\rho}^k_{AB}) +S(\{\bar{p}_k\})=2nE^C_{\rm sq}(\rhoAB)+S(\{\bar{p}_k\})$. The second inequality in \eqref{eq:sequence} is obtained by the standard regularization of both sides of $2E^C_{\rm sq}\geq I^{(\infty)}_b$.
%and weak additivity of $I^{(\infty)}_b$.
\end{proof}

Theorem \ref{thm:sequence} together with Lemma \ref{lem:strict}, provides a new proof that the entanglement cost $E_c$ --- the rate at which one has to consume pure entanglement to create many copies of a given state via LOCC --- is strictly positive for all entangled states~\cite{dong2005}. Indeed, by using the formula for $E_c$ of ~\cite{entcost}, one easily checks that $E_c\geq\big( E^C_{\rm sq}\big)^{(\infty)}$.

Finally, we notice that one may define a variant of $I^{(\infty)}_b$
%by imposing restrictions on the allowed $n$-copy broadcast states. One interesting
%restriction consists
by considering broadcast copies only in the class of permutation-invariant states, that is, states $\rho^{(n)}_{A^nB^n}$ satisfying $\rho^{(n)}_{A^nB^n}=\pi\rho^{(n)}_{A^nB^n}\pi^{-1}$, for all permutations $\pi$ of the $n$ pairs $A_iB_i$. Thus, we define the \emph{symmetrical broadcast regularization of MI} as:
\[
I^{(\infty)}_{b,{\rm sym}}(\rho_{AB})\equiv \lim_n \frac{1}{n}\min_{\rho^{(n)} \ \text{perm-inv}}I(A^n:B^n)_{\rho^{(n)}}.
\]
For such a quantity we are able to establish the following.
\begin{theorem}
\label{thm:sim}
$I^{(\infty)}_{b,\rm sym}=2E^C_{\rm sq}$, i.e., symmetric copies of the form $\rho^{(n)}[\{(p_k,\rho^k_{AB})\}]$ are asymptotically optimal.
\end{theorem}
This theorem can be interpreted as support for our conjecture, since it implies that the symmetric version of $I^{(\infty)}_{b}$ is an entanglement monotone.

By Theorem~\ref{thm:sequence} it suffices to prove the direction ``$\geq$''. The intuition is that permutation-invariant states can be approximated by de Finetti states~\cite{deFin, definettirenner}.
%by means of the quantum de Finetti theorem
This idea can be made precise with the help of the so-called exponential de Finetti theorem, in particular by showing that entropy is ``robust'' under the disturbance of a small number of subsystems~\cite{definettirenner} (see~\cite{longAmutual information} for details).
\begin{lemma}
\label{lem:entsym}
 Let $\rho_{A^nB^n}$ be a permutation-invariant state on $(\cH_A \otimes \cH_B)^{\otimes n}$. Then there exists an ensemble of states $\{ p_i, \rho_{AB}^i\} $, where $\rho_{AB}^i$ are states on $\cH_A\otimes \cH_B$, such that $S(X^n)_{\rho_{A^nB^n}}=n\sum_i p_i  S(X)_{\rho^i_{AB}}+o(n)$, for $X=A,B,AB$, and $\|\rhoAB-\sum_i p_i \rho_{AB}^i\|_1=o(1)$, with the reduced state $\rhoAB=\rho_{A_kB_k}$, $k=1,\ldots,n$.
\end{lemma}
Theorem~\ref{thm:sim} follows because for any permutation-invariant broadcast copy $\rho^{(n)}$ of $\rho$, the continuity of $E^C_{\rm sq}$~\cite{multisquash} and Lemma~\ref{lem:entsym} assure the existence of an ensemble $\{(p_i, \sigma^{AB}_i)\}$ such that $I(\rho^{(n)}[\{(p_i, \sigma^{AB}_i)\}])\leq I(\rho^{(n)})+o(n)$.

%\begin{proof}%[Proof of Theorem~\ref{thm:sym}]
%Since $\rho^{(n)}[\{(p_k,\rho^k_{AB})\}]$ is symmetric, one can argue as in the proof of Theorem~\ref{thm:sequence} that $I^{(\infty)}_{b,\rm sym}\leq 2E^C_{\rm sq}$.
%\end{proof}

%\medskip
%\noindent
%\emph{Conclusions---}
\section{Conclusions}
In this Letter, we have introduced a new way of quantifying the \emph{quantumness} of correlations. This led us to define a new correlation measure, the broadcast regularization of mutual information. Its strict positivity on and only on entangled states can be interpreted as a signature of the \emph{monogamy of entanglement} for \emph{any} entangled state. Our study furthermore reveals a novel relation between extensions ---here broadcast extensions--- and entanglement, a topic of practical interest~\cite{extQKD}.

Focus has been on correlations between two parties. As in~\cite{localbroad}, our results can be straightforwardly extended to the multipartite case if a suitable definition of multipartite mutual information is adopted.

%The standard notion of monogamy of entanglement refers to the impossibility for a system to be strongly entangled a the same time with two or more other systems, while here we find a copy-copy monogamy of total correlations in the broadcast scenario. Correlations involving entanglement, contrary to the classical correlations of CC-states and ---up to a certain approximation--- separable states, cannot be freely shared among copies.
%: each broadcast copy needs its own amount of correlations.

We thank M. Horodecki, R. Renner, and B. Toner for discussions. This work began when MP was a Lise Meitner Fellow
%at the Institute for Theoretical Physics
at the University of Innsbruck and is partially supported by the Austrian Science Fund (FWF). We acknowledge support: by
%IQC
QuantumWorks and Ontario Centres of Excellence [MP and CM];  by the Excellence Network of Bavaria (TMP, QCCC)
and the SFB 1388 of the German Science Foundation [MC]; by EU IP SCALA and the LFPPI network [PH].

%Suppose that Alice and Bob each possess two systems ($AA'$ and $BB'$, respectively), and let the global quantum state of the four subsystems be described by a density matrix $\rho_{AA'BB'}$. Alice and Bob can extract classical correlations from the shared quantum system via local measurements. We show that for Alice and Bob to be sure that they can extract more classical correlations from $\rho_{AA'BB'}$ than from the subsystems $A'B'$ alone it is necessary and sufficient that the reduced state $\rho_{AB}$ be entangled.

%\renewcommand{\refname}{\vspace{-1cm}}

\end{document}